\documentclass[aps,prl,twocolumn,showpacs,floatfix,superscriptaddress]{revtex4}
\usepackage{epsfig,graphicx}

\begin{document}

\title{
Modulation of the local density of states within the $d$-density wave theory in the underdoped
cuprates
}

\author{Amit Ghosal}

\affiliation{Department of Physics, Duke University,
Durham, North Carolina 27708-0305}

\author{Angela Kopp}
\affiliation{Department of Physics, University of California Los Angeles,
Los Angeles, California 90095-1547}

\author{Sudip Chakravarty}

\affiliation{Department of Physics, University of California Los Angeles,
Los Angeles, California 90095-1547}

\date{\today}

\begin{abstract}
The low temperature scanning tunneling microscopy spectra in the underdoped
regime is analyzed from the perspective of coexisting $d$-density wave (DDW) and $d$-wave
superconducting states (DSC). The calculations are carried out in the presence of a
low concentration of unitary impurities and within the framework of the fully
self-consistent Bogoliubov-de Gennes theory, which allows local modulations of the magnitude of the order parameters
 in response to the impurities. Our theory captures  the
essential aspects of the experiments in the underdoped BSCCO at very low temperatures.

\end{abstract}


\maketitle
A fundamental tension in the field of high temperature superconductors is the notion of a competing order parameter in the underdoped regime, which can provide a natural explanation of why the superconducting dome exists and shed light on the nature of the pseudogap. While a charge ordered state is a candidate \cite{c1}, one of us has proposed, and extensively studied, a new order parameter, which results in circulating currents arranged in a staggered pattern(DDW) \cite{ddw}. Many experiments are consistent with this order parameter, as demonstrated in studies of the superfluid density, the polarized neutron scattering, the Hall number in pulsed 60 T magnetic field, the angle resolved photoemission spectroscopy (ARPES), the lack of specific heat anomaly at the pseudogap temperature, the transition temperature in multilayer cuprates, and the infrared Hall angle measurements \cite{otherexpddw}. So far the clinching direct experiment, the polarized neutron scattering, has remained suggestive \cite{Mook} because of the difficulty of detecting weak signals from the small magnetic moments generated by the circulating orbital currents. Recently, another novel experimental test has been suggested that takes advantage of the spin-orbit coupling in the DDW state \cite{Wu}.

Here we turn to the intriguing scanning tunneling
microscopy (STM)
measurements \cite{hoffman,howald,mcelroy1,mcelroy2,yazdani,kapitulnik}. In spite of numerous theoretical analyses of this problem \cite{Scalapino,Scalapino2,Lee,bena,bena2,
c1,c2,c3,c4,zhang,c6,franz},  no
comprehensive  theoretical picture has yet emerged, although certain aspects are
captured by some of them. For example, earlier measurements in the 
slightly overdoped samples at very low temperatures have been
elegantly explained  in terms of a quasiparticle scattering interference model, named the
octet model \cite{mcelroy1,Lee}. At the same time, an interpretation in terms of dynamic charge
fluctuations has also been advanced \cite{c1,kapitulnik}. The focus here, however, is on an extensive 
set of experiments as a function of doping in BSCCO at very low temperatures \cite{mcelroy2}. The exciting finding 
of these experiments is the emergence of a new order, present along with the $d$-wave 
superconductivity (DSC). The salient signature  is a sudden development of a relatively 
non-dispersive incommensurate wave vector, $q^{*}$, at higher energies in the underdoped 
regime. 

In this Letter we explain the experiments by adopting a view that is orthogonal  to the notion of charge order
and consider DDW.  At first sight, this would seem impossible, as STM is not sensitive to currents.
The key to this puzzle lies in  the presence of disorder (inevitable in these materials)  that leads to variations
of the charge density, which in turn can scatter the quasiparticles of the DDW, revealing indirectly the DDW order in the form of an interference pattern. The idea is not as
surprising as it may seem, because, after all, this was precisely the basis of  the successful octet model.
Clearly,  DSC corresponds to off-diagonal long range
order rather than diagonal long range order and therefore cannot directly affect STM. The general message is that a large class of ordered states can leave their characteristic signatures via the scattering interference of their distinctive quasiparticle spectra. Our analysis is broadly consistent with the conclusions drawn from the recent comparisons of the autocorrelation function of the ARPES spectra and the STM spectra \cite{AC1,AC2}.

A single-impurity $T$-matrix
calculation in the coexisting DDW and DSC state \cite{bena,bena2} was unable to recover  the salient feature of the experiments, while pin-pointing certain important aspects . We shall show that the fault lies with the method and not intrinsically with the idea that  a new order arises in  the low temperature underdoped regime, namely the coexisting DDW  and DSC order. The problem is that the $T$-matrix approach neither allows the amplitude of the
order parameter of the DSC to modulate in response to the impurity potential
nor allows a proper treatment  of the current conservation in the DDW state
in the presence of impurities \cite{ghosal1,ghosal04}.  Moreover, it 
excludes not only spatial structures but also the interference of the impurities.

To understand the effect of various competing orders on the local density of states (LDOS), all we need to know are  the corresponding wave functions in the presence of finite concentration of impurities. We accomplish this by choosing 
the simplest Hamiltonian that yields both DSC and DDW, and coexisting
DDW and DSC, states within the Bogoliubov-de Gennes mean field theory. We use this Hamiltonian solely as a crutch to generate the states with the broken symmetries that we wish to study.
In general, deep within the superconducting dome, that is, at very low temperatures, there is good evidence
of the existence of quasiparticles. The experiments 
that we address here are precisely in the regime in which our theory is expected to be valid. The corollary to these observations is that a straightforward finite temperature extension of our theory cannot be applied to the experiments of Ref.~\cite{yazdani} because the simple picture of quasiparticles may  not be valid at higher temperatures and above the superconducting dome. 

The Hamiltonian is
${\cal H} = {\cal K} + {\cal H}_{\rm int} + {\cal H}_{\rm dis}$,
where 
${\cal K} = -\sum_{ij,\alpha} (t_{ij} c_{i\alpha}^{\dag} c_{j\alpha}+ h.c.)$.
The operator $c^{\dag}_{i\alpha}$ ($c_{i\alpha}$) creates (destroys) an electron
at the site $i$ with spin $\alpha$, and $t_{ij}$ is the hopping matrix element
to the nearest ($t$) or to the next-nearest ($-0.273t$) neighbor \cite{norman}.
The interacting part consists of 
\begin{displaymath}
{\cal H}_{\rm int} =
J\sum_{<ij>}\left({\bf S}_i \cdot {\bf S}_j - \frac{n_i n_j} {4} \right)
- W\sum_{<ij>,\alpha,\beta} n_{i \alpha} n_{j \beta}, 
\end{displaymath}
which is identified
by two parameters $J$ and $W$~\cite{fnote1} (${\bf S}_{i}$ is the spin operator at the site $i$ and $n_{i \alpha}$ is the density operator).
Disorder is introduced through
${\cal H}_{\rm dis} = \sum_{i\alpha} \left(V_i-\mu \right) n_{i\alpha}$,
$V_i$, at each site $i$, is an independent random variable, which is either $+V_0$
(repulsive) with a probability $n_{\rm imp}$ (impurity concentration) or
zero, and $\mu$ is the chemical potential of the system.

We solve the above model within the self-consistent framework of
BdG mean field theory, which amounts to decoupling
the interaction terms, resulting in  a ${\cal H}_{\rm eff}$ that leads to the
{\it local} order parameters:  DSC order ($\Delta_i$), DDW order
($\chi_i$) and also the local Hartree and Fock shifts.  The
details can be found in Ref. ~\cite{ghosal04}.  The chemical potential $\mu$ is obtained by fixing the average
density of electrons, $\langle n \rangle=\sum_i \langle n_i \rangle /N$,
at the desired value ($N$ being the system size).

We concentrate  on the results at $T=0$ for parameters
$J=1.6$,
$W=0.6$ in units of $t$ (all energies will be expressed in this
unit and all lengths will be expressed in tems of the lattice spacing),
and $\langle n \rangle=0.9$ (equivalently 10\% hole doping) on square
lattices with a unit cell of size $N=30 \times 30$. Larger unit cells result in negligible improvements at the cost of computer time; the dimensionality of the matrices for which all the eigenvalues and the eigenvectors must be repeatedly obtained is $2N\times 2N$, that is, $1800\times 1800$. These parameters result
in $\chi_0=0.33$ and $\Delta_0=0.12$ in a disorder free system in the
coexisting DDW and DSC phases, where $\chi_0$ is defined through
$\chi_{\bf k}=\chi_0 (\cos k_x- \cos k_y)/2$ and similarly for $\Delta_0$.
Results for systems with solely DDW or DSC order will also be
discussed for slightly different parameters: $J=1.4, W=0.5$.
We choose $V_0=100$, close to the unitary limit, and results are averaged
over 8-10 different realizations (8 for the DDW and the DSC orders by themselves, but 10 for the coexisting DDW and DSC) of the random potential. For better
statistics we use the repeated zone scheme ~\cite{ghosal04,RZS} in
which a super-cell is constructed by replicating the $30\times30$ unit cell 10 times along the $x$-direction and 10 times along the $y$-direction. This provides a denser set of energy eigenvalues.

\begin{figure}[htb]
\includegraphics[width=3.25in,clip]{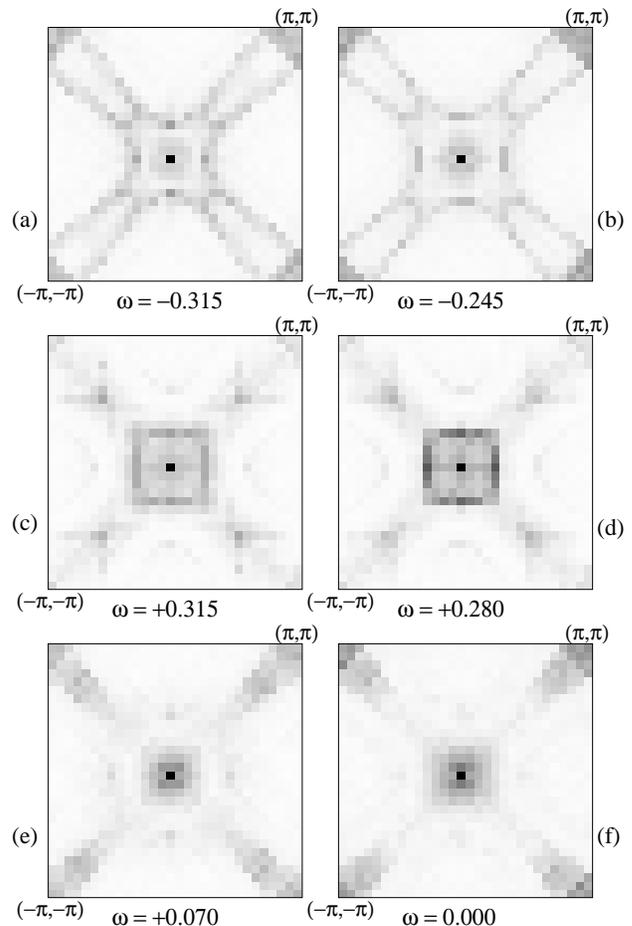}
\caption{
Density plots on a linear scale of the FT-LDOS within the first Brillouin zone in the $k_{x}-k_{y}$ plane. The panels (a) through (f) correspond to different energies, $\omega$, for the coexisting DDW+DSC phase. Here  $n_{\rm imp}=0.01$ and $V_0=100$.
Peaks at $q^* \sim 0.15 (2\pi)$ are clear in panels (a) through (d). At
smaller energies, panels (e) and (f), $q^*$-peaks are absent and a broad peak
appears near ${\bf q}=0$, whose width  decreases with increasing $|\omega|$. The strength of the peak at ($0,0$) is reduced for the clarity.
}
\label{fig:Fig1}
\end{figure}

In the presence of impurities, both orders become spatially inhomogeneous.
However, at $n_{\rm imp}=0.01$, the (disorder) averaged DDW order degrades
much less ($\overline{\chi_i} \approx 0.95\chi_0$) compared to the DSC order
($\overline {\Delta_i } \approx 0.55\Delta_0$), consistent with earlier
findings~\cite{ghosal04}.
We define the  Fourier transform $N({\bf q},\omega)=\sum_{\bf r} e^{-i{\bf q.r}}
 N({\bf r},\omega)$, where $N({\bf r},\omega)$ is the LDOS at a site ${\bf r}$
with energy $\omega$. The disorder averaged power spectrum $P({\bf q},\omega)=\overline{|N({\bf q},\omega)|^{2}/N}$
is calculated and then $\sqrt{P({\bf q},\omega)}$ is compared with experiments, as this is the theoretical measure \cite{Scalapino2} of the  Fourier transform of the experimentally measured $dI(V,{\bf r})/dV$, where $I$ is the tunneling current, and  $V$ is the bias voltage ($\omega=eV$).

The results  in the coexisting DDW+DSC state
are shown in  in Fig.~\ref{fig:Fig1}. The intensity of the peaks along ($0,\pm 1$) and ($\pm 1,0$)
appears as a generic and robust
property, more precisely at  $q^* \sim (0.15) 2\pi$ for the chosen set of parameters. 
The $q^{*}$ peaks occur for $|\omega| > \Delta_0$.
At lower energies the FT-LDOS profile looses periodic $q^{*}$ modulations
and a rather broad peak occurs in its Fourier transform
at ${\bf q} \approx 0$ (See Figs. 1(e) and 1(f)).
Unitary impurity resonances~\cite{davis} dominate this regime (near $\omega
\approx 0$) and wash out the globally periodic LDOS modulations arising from
interference. Unitary scatterers do not affect the spectrum at energies larger
than $\Delta_0$, and hence the LDOS modulations are preserved.

The finite system, $30 \times 30$, allows a {\bf q}-resolution
of $0.033$, which is rather coarse compared to the real data (the super-cell
does not introduce any additional independent wave vectors due to the
periodicity). Along with the  strong peak at ${\bf q} = 0$  mentioned above,
the resolution problem prevents us from observing the low energy modulation
$q_{1}$, in the terminology of Ref.~\cite{mcelroy2};
it is hidden by the broad peak
around ${\bf q} = 0$, whose width decreases with
$|\omega|$.
On the other hand,  $q_{5}$ is quite visible (at least
for $\omega > 0$). It is known from Ref.~\cite{mcelroy2} that the intensities
of $q_{2}$, $q_{3}$, $q_{6}$, $q_{7}$ relative to $q_{1}$ and $q_{5}$ fall
with decreasing doping and $q_{4}$ is not seen at all \cite{AC1,AC2}. This is consistent with our
calculation, which was performed in the underdoped regime.

\begin{figure}[htb]
\includegraphics[width=3.25in,clip]{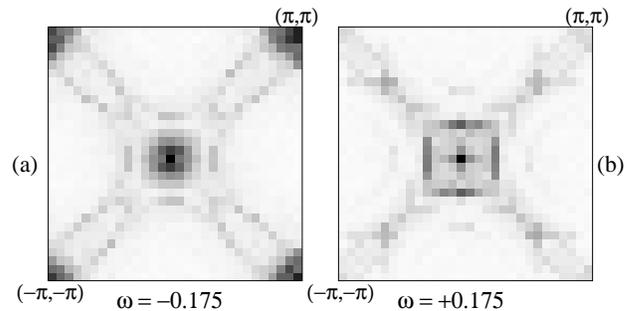}
\caption{
FT-LDOS as in Fig.1 but for DDW
($\chi_0 \approx 0.28$). 
DSC order is forced to zero on all sites. Qualitative features
are similar to Fig.1 for a wide range of $\omega$.
}
\label{fig:Fig2}
\end{figure}

\begin{figure}[htb]
\includegraphics[scale=0.3]{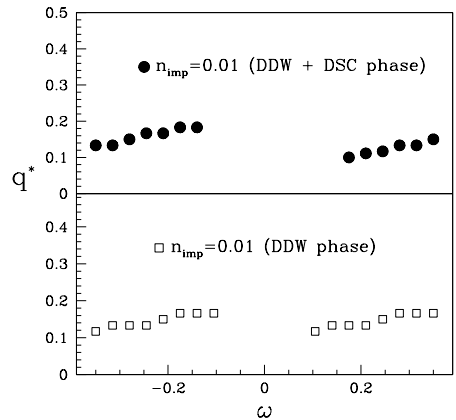}
\caption{
Top panel: dispersion of $q^*$ (in 
units of  $2\pi$) with $\omega$ in an impure DDW+DSC phase.
Bottom panel: the same as in top panel but for the  DDW phase.  The error bars are about the size of the symbols.
}
\label{fig:Fig3}
\end{figure}

Because DSC order is already weak in the pure system for our chosen parameters,
and it rapidly gets weaker with impurities, it is important to
study profiles similar to Fig.1 in the DDW phase {\it without} coexisting
DSC order with $n_{\rm imp}=0.01$. The results are 
presented in Fig.2.  The features are similar to those
of Fig.1, but the $q^{*}$ peaks first emerge for a little smaller value of $|\omega|$ (see, Fig.~\ref{fig:Fig3}) . The
spectra shown in Fig.~\ref{fig:Fig2} are generic, and  similar results
are obtained  for $n_{\rm imp}=0.02$.  We have also studied our model in the
DSC-only state in the same underdoped region and have not found $q^{*}$ peaks. 

An intuitive explanation of $q^{*}$ is as follows. It is due to scattering between the tips of the hole pockets and the sharp rise of the pseudogap. In the coexisting DDW and DSC state, the low-energy constant energy contours are ``bananas'', essentially from the pure DSC order, except that they are doubled up because of the broken translational symmetry. However, the disorder considerably smears out these scattering events. As the energy increases these bananas coalesce and,  then on, the scattering is between the tips of the hole pockets of DDW. Since DDW is less affected by potential scattering than DSC, the signature becomes robust.

The dispersion of the $q^*$-peaks is shown in Fig.~\ref{fig:Fig3}. 
The dispersion in the top panel of  Fig.~\ref{fig:Fig3} is somewhat stronger than in
Ref.~\cite{mcelroy2} but gets weaker in the  DDW-only  phase (bottom panel). Similar
results are also found for $n_{\rm imp}=0.02$. The value of
$q^* \sim (0.15)2\pi$  is due to the chosen set  of band parameters, which
will change with a different choice.
\begin{figure}[htb]
\includegraphics[scale=0.3]{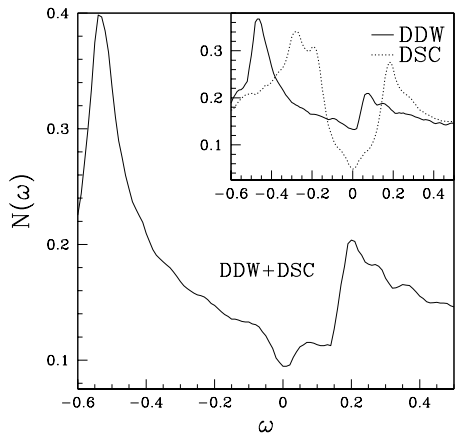}
\caption{
Impurity averaged normalized DOS, $N(\omega)$, for  coexisting DDW+DSC
for $n_{\rm imp}=0.01$. The spectrum is
asymmetric and low energy
humps appear around $\Delta_0$. The inset corresponds to
DDW ($\chi_0 \approx 0.28$) and DSC ($\Delta_0 \approx 0.196$), with $n_{\rm imp}=0.01$,  $J=1.4$, and $W=0.5$.
}
\label{fig:Fig4}
\end{figure}

We have repeated our calculations with impurity strengths $V_{0}=10 $ and
$V_{0}=1$. The results are essentially the same for $V_{0}=10 $, while for
$V_{0}=1$ (non-unitary scatterers) the spectra for negative energy are
somewhat different, though the key features are preserved. Thus the robustness
of the results found here is heartening.
We emphasize, however, that the effects of inhomogeneity and impurity interactions
can be very subtle~\cite{hirschfeld}.

The disorder {\it averaged} DOS, $N(\omega)$,
is shown in Fig.~\ref{fig:Fig4}.
We see that $N(\omega)$ is asymmetric in the DDW phase, both with and without
DSC order. The asymmetry in the spectrum has the same origin as the asymmetries in Figs.~\ref{fig:Fig1} and \ref{fig:Fig2}. In the coexisting phase, the remnants of the  superconducting
density of states peaks can be seen, below which modulation of LDOS dies out.
The spectrum is  tantalizingly similar to the experimental results of Ref.~\cite{mcelroy2}.

To summarize, the inhomogeneous phase of coexisting
DDW and DSC captures  the essential experimental 
features seen in the  underdoped regime at low temperatures, namely the 
emergence and the nature of the relatively nondispersive $q^{*}$ peaks seen only at higher energies;
in contrast, the DSC order alone is inadequate. From our perspective the emergence of $q^{*}$ is not due to an explicit charge order, fluctuating or otherwise, but due to the current modulations of the DDW.  It would be interesting to see if $q^{*}$ extends to lower energies with purer samples. The interpretation of the experiment \cite{Hanaguri} in $\mathrm{Ca_{2-x}Na_{x}CuO_{2}Cl_{2}}$ is an open issue.

This work was supported by the grant NSF-DMR-0411931. We would like to
thank C. Bena, J. C. Davis, J. Hoffman, H. -Y. Kee, K. McElroy, S. Kivelson, and D. J. Scalapino  for helpful comments.
We acknowledge the computational support of H. U. Baranger's group at the Duke University, where
part of the calculations were carried out.


\end{document}